\documentstyle[twoside,fleqn,espcrc2,psfig]{article}

\newcommand{\AmS}{{\protect\the\textfont2
  A\kern-.1667em\lower.5ex\hbox{M}\kern-.125emS}}
\newcommand{\be}{\begin{equation}}
\newcommand{\ee}{\end{equation}}
\newcommand{\ben}{\begin{eqnarray}}
\newcommand{\een}{\end{eqnarray}}
\newcommand{\nn}{\nonumber}
\newcommand{\slas}[2]{{{#1}\hspace{-5pt}{/}}_{#2}}
\newcommand{\slal}[2]{{{#1}\hspace{-5pt}{/}}_{#2}}
\def\simgt{\rlap{\lower 3.5 pt\hbox{$\mathchar \sim$}}\raise 1pt \hbox {$>$}}
\def\simlt{\rlap{\lower 3.5 pt\hbox{$\mathchar \sim$}}\raise 1pt \hbox {$<$}}

\title{Light Hadron Weak Matrix Elements}

\author{Yoshinobu Kuramashi
        \address{Department of Physics, Washington University, 
        St.~Louis, Missouri 63130}
\thanks{On leave from Institute of Particle and Nuclear Studies,
High Energy Accelerator Research Organization(KEK),
Tsukuba, Ibaraki 305-0801, Japan}}
       
\begin{document}

\begin{abstract}

Recent developments in lattice QCD calculation
of weak matrix elements involving light quarks are described.
We focus on four topics: $B_K$ with the Kogut-Susskind and Wilson 
quark actions, $\Delta I=1/2$ rule, proton decay matrix elements
and the application of domain wall QCD
to calculation of weak matrix elements. 

\end{abstract}

\maketitle

\section{Introduction}

 
The kaon bag parameter $B_K$, which is required
for constraining the CP violation parameter in the
Cabibbo-Kobayashi-Maskawa matrix from experiment, 
has been successfully calculated with the Kogut-Susskind(KS) 
quark action, taking advantage of its ${\rm U_A}(1)$ 
chiral symmetry\cite{bk_ks_init}. 
A recent systematic calculation 
of $B_K$ with high statistics\cite{bk_ks_jlqcd}, however,
reveals that the systematic uncertainties in the perturbative 
renormalization 
factors to connect lattice and continuum operators 
have sizable magnitude, which indicates the necessity of
non-perturbative renormalization. 
For the calculation of $B_K$ with the Wilson quark action
we have to deal with the operator mixing problem caused by
the explicit chiral symmetry breaking in the action,
which 
turned out to be not 
effectively treated by 
perturbation theory\cite{bk_w_bs}.
Several non-perturbative methods have been proposed
to control the operator mixing. 

The CP conserving $K\rightarrow \pi\pi$ decay amplitudes
give an intriguing testing ground for lattice QCD calculations.
It would be exciting to be able to show that QCD explains
the $\Delta I=1/2$ rule: $A_0/A_2\approx 22$ where 
$A_{0,2}=A(K\rightarrow \pi\pi[I=0,2])$.
Although studies of $K\rightarrow \pi\pi$ decay amplitudes 
with lattice QCD were initiated a decade ago,
little progress has been achieved since then.
Recent work on the one-loop calculation of $K\rightarrow \pi\pi$ 
decay amplitudes using chiral perturbation 
theory(ChPT)\cite{kpipi_gl,kpipi_gp}, 
however, provides some important information 
about the lattice results.

In lattice QCD studies for the proton decay amplitude
$\langle \pi^0 | {\cal O}^{\slal{B}{}} |p\rangle$,
pioneering work estimated it
from the matrix element $\langle 0 | {\cal O}^{\slal{B}{}} |p\rangle$
with the aid of the chiral lagrangian\cite{pd_hara,pd_bowler}, which
were followed by the direct measurement of 
$\langle \pi^0 | {\cal O}^{\slal{B}{}} |p\rangle$\cite{pd_gavela}.  
The results showed an unexpectedly large discrepancy
between these two methods:
the direct method yielded a two or three times smaller value 
than the indirect one.
This peculiar feature was confirmed by JLQCD last year\cite{pd_jlqcd_98}.
At this conference, however, JLQCD\cite{pd_jlqcd_99} pointed out that
the calculational method in refs.\cite{pd_gavela,pd_jlqcd_98}
was wrong, and presented the correct estimate of 
the proton decay amplitudes with the direct method.
    
The domain wall quark formulation in lattice QCD\cite{dwf_shamir,dwf_fs}
potentially has superior features 
over the Wilson and KS quark actions:
realization of the chiral limit on arbitrary number of flavors
without fine tuning. Recent simulation results for $B_K$ show
a good chiral behavior\cite{dwf_blum}, which encourages us to
apply the domain wall quark to the calculation of weak matrix
elements. Some progress has been made this year numerically and 
theoretically.

This article is organized as follows. In Sec.~2 we present
the $B_K$ results with the KS and Wilson quark actions.
From the examination of the systematic errors in the KS result
we explain the reason why the non-perturbative renormalization is
necessary. The present status of the non-perturbative 
renormalization method
is also discussed. Results for the  $K\rightarrow \pi\pi$ 
decay amplitudes are presented in Sec.~3, where we focus on the
one-loop effects of ChPT. In Sec.~4 we describe an essential advance on 
the calculation of the proton decay matrix elements.
This year's progress on the calculation of weak matrix elements
with the domain wall QCD is explained in Sec.~5.
Our conclusions are summarized in Sec.~6. 
Results for structure functions have been reviewed by 
Petronzio\cite{structuref}.

\section{$B_K$}

\subsection{$B_K$ with the KS quark action} 

The $K$ meson $B$ parameter is defined by
\be
B_K=\frac{\langle
{\bar K}^0 \vert \bar{s}\gamma_\mu(1-\gamma_5)d\cdot
\bar{s}\gamma_\mu(1-\gamma_5)d \vert K^0 \rangle}
{\frac{8}{3}\langle
{\bar K}^0 \vert \bar{s}\gamma_\mu\gamma_5 d\vert 0\rangle
\langle 0 \vert\bar{s}\gamma_\mu\gamma_5 d \vert K^0 \rangle},
\label{eq:bk_def}
\ee
for which we quote results in the naive dimensional
regularization(NDR) scheme at the renormalization scale 2GeV.
In the ratio of eq.(\ref{eq:bk_def}) the fluctuations are largely canceled
between the numerator and the denominator, 
which enable us to easily achieve high statistical accuracy.  
Taking advantage of this virtue, we have 
had control over the systematic errors step by step.
As of this conference $B_K$ is the weak matrix element with which
we have had the most success. 

\begin{figure}[t]
\centering{
\hskip -0.0cm
\psfig{file=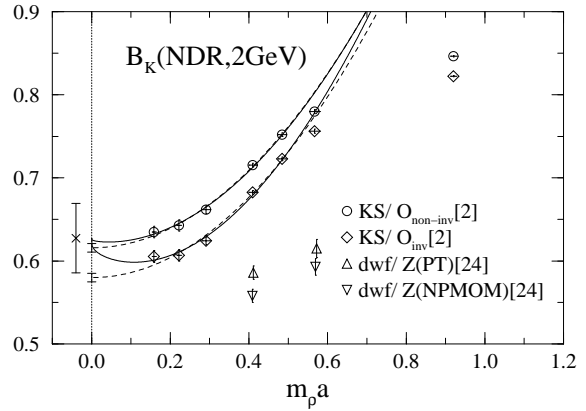,width=75mm,angle=-90}
\vskip -10mm  }
\caption{Quenched $B_K({\rm NDR, 2GeV})$ with the KS quark action 
as a function of $m_\rho a$.
See text for the solid and dashed lines. Domain wall fermion results
will be explained in Sec.~\protect{\ref{sec:dwf}}.} 
\label{fig:bk_ks}
\vspace{-8mm}
\end{figure}

Figure~\ref{fig:bk_ks} presents a recent result
for a systematic quenched $B_K$ calculation performed by
JLQCD  employing seven values of lattice spacing 
with $m_d=m_s$\cite{bk_ks_jlqcd}.
The non-local weak four-fermi operator is constructed
either in gauge invariant way with insertion 
of the link variables (${\cal O}_{\rm inv}$)
or in non-invariant way without link variables in Landau gauge
(${\cal O}_{\rm non-inv}$).
Each operator is perturbatively matched to the operator 
in the continuum NDR scheme
at the scale $q^* =1/a$ employing tadpole improvement.
The value of $B_K$ at $\mu=2$GeV is obtained via a two-loop
renormalization group running from $\mu=1/a$GeV.

\begin{figure}[t]
\centering{
\hskip -0.0cm
\psfig{file=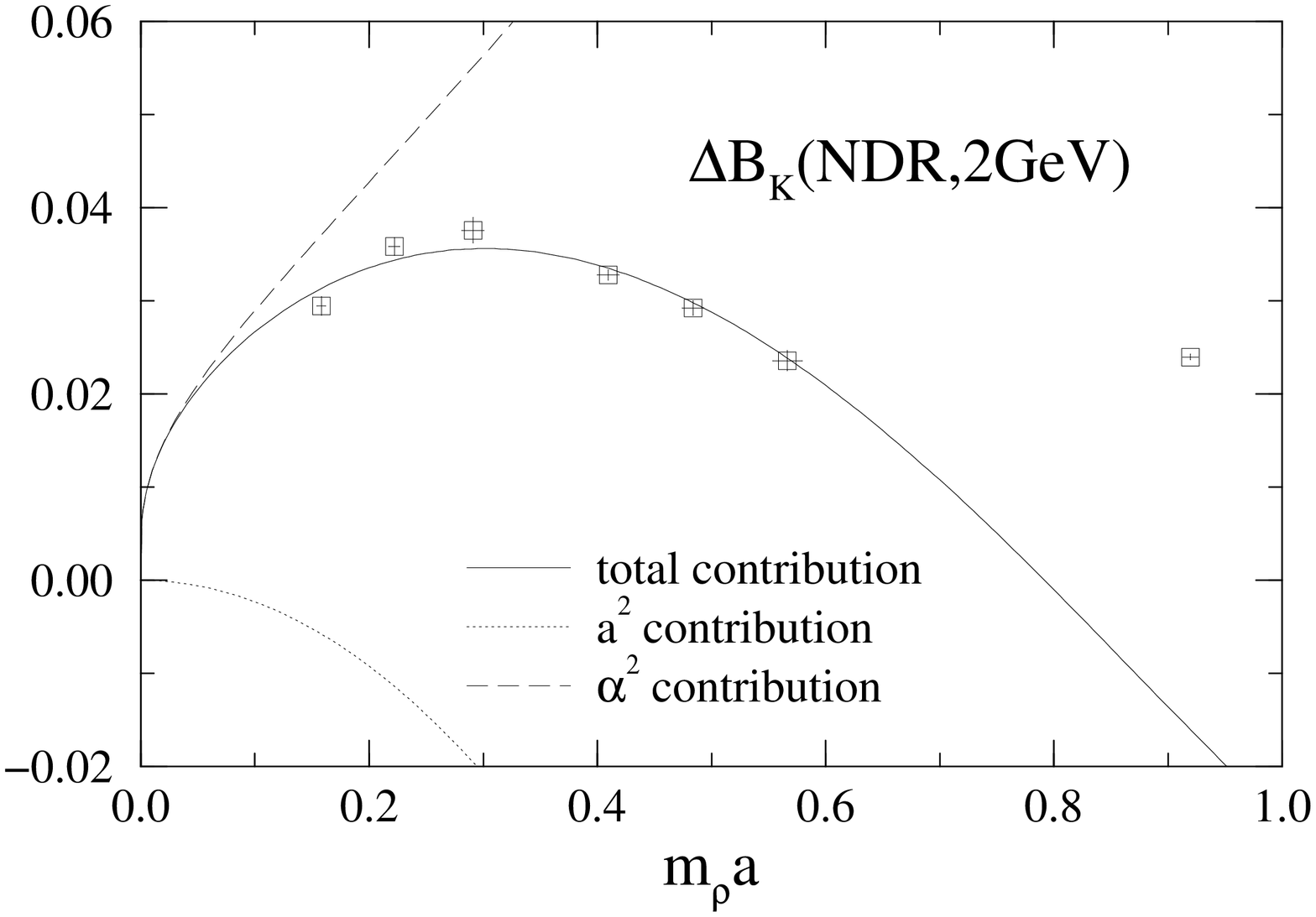,width=75mm,angle=-90}
\vskip -10mm  }
\caption{ Difference of $B_K({\rm NDR, 2GeV})$ 
between ${\cal O}_{\rm inv}$ and ${\cal O}_{\rm non-inv}$
as a function of $m_\rho a$.} 
\label{fig:bk_ks_diff}
\vspace{-8mm}
\end{figure}

The statistical error is $0.1\%$ at the largest
lattice spacing and gradually increasing to $1.2\%$ 
at the smallest one. Let us examine
conceivable systematic errors in order.
(i) Quark mass dependence:
$B_K$ at the physical kaon point is accessible
with interpolation, not extrapolation, which means
the interpolated value is strongly robust against the choice of
fitting function.
(ii) Finite size effects: 
Finite size studies at $\beta=6.0$ and $6.4$ show that
the magnitude of the size dependence decreases to less than 
$0.5\%$ for $L\geq 2.2$fm. The spatial size is kept larger than
$L\simeq 2.3$fm at all $\beta$.
(iii) Scaling violation:
The five data points below $m_\rho a\simeq 0.6$ for each operator
are consistent with $O(a^2)$ scaling behavior 
expected theoretically\cite{bk_ks_a^2}.
A quadratic fit of the five points(dashed lines in Fig~\ref{fig:bk_ks})
gives a value at the continuum limit $B_K=0.616(5)$ for
${\cal O}_{\rm non-inv}$ and $0.580(5)$ for ${\cal O}_{\rm inv}$.
This $7\sigma$ discrepancy raises the question of what causes this
difference.
(iv) $O(\alpha^2)$ uncertainties:
As seen in the matching procedure of ${\cal O}_{\rm inv,non-inv}$
to the continuum operator, their difference $\Delta B_K$ 
should be of not only $O(a^2)$ but also $O(\alpha^2)$.
Figure~\ref{fig:bk_ks_diff} illustrate that the difference
is well described by the function 
$\Delta B_K=b_1(m_\rho a)^2+b_2(\alpha_{\overline {\rm MS}}(q^*))^2$
below $m_\rho a\simeq 0.6$, which suggests that
a $\alpha_{\overline {\rm MS}}^2$ term should be incorporated
in the fit of $B_K$.
An attempt to simultaneously fit $B_K$ for both operators
is shown in Fig.~\ref{fig:bk_ks}(solid lines), which
yields $B_K=0.628(42)$ in the continuum limit(cross). This $7\%$ error,
which is roughly 10 times larger than those with naive
quadratic fit, reflects the magnitude of the $O(\alpha^2)$ 
uncertainties in the operator matching procedure. 
To eliminate the $O(\alpha^2)$ uncertainties,
it would be necessary to adopt the non-perturbative
renormalization method. 
(v) Quenching effects:
In 1996 the OSU group studied the dynamical quark effects
on $B_K$ employing three ensembles of gauge configurations at 
$a^{-1}\approx 2$GeV:
$N_f=0$ at $\beta=6.05$, $N_f=2$ at $\beta=5.7$ 
and $N_f=4$ at $\beta=5.4$\cite{bk_ks_dq}. 
They found that $B_K$ with $N_f=3$ is $5\pm 2\%$ larger than the
quenched value.
However, it is hard to assume this number as the dynamical quark effects
in the continuum limit: we cannot exclude the possibility
that the scaling violation of $B_K$ in full QCD might be different from
the quenched case.
This fact compels us to perform the systematic calculation of $B_K$
in full QCD varying the lattice spacing
and the dynamical quark mass.  
(vi) Degenerate quark mass $m_s=m_d$:
For $B_K$ in the quenched QCD
there is a divergent chiral logarithm
as $m_d\rightarrow 0$ in the chiral expansion 
of ChPT\cite{bk_ks_chpt_nd,kpipi_gl}.
We can investigate the $m_s\neq m_d$ case only in the full QCD.

The next step toward the precise determination of $B_K$  
is the control of the $O(\alpha^2)$ uncertainties.
We also note that the systematic full QCD calculation of $B_K$ with
$m_s\neq m_d$ is now feasible in view of the performance of 
the current full QCD simulations\cite{fullqcd_cppacs}.

\subsection{Non-perturbative renormalization}

A naive non-perturbative renormalization method is to
copy what is done in perturbative renormalization
in a non-perturbative way\cite{npmom}.
The renormalization condition in the momentum-space subtraction scheme
(MOM) for a certain operator
is provided by imposing that the off-shell quark matrix element
in a fixed gauge should coincide with their tree level value.
For example, if we consider the bilinear operator
${\cal O}_\Gamma={\bar\psi}\Gamma\psi$,
the condition is given by
$
Z_\Gamma(g_0,\mu a)
\langle q(p)|{\cal O}_\Gamma|q(p)\rangle|_{p^2=\mu^2}
=\langle q(p)|{\cal O}_\Gamma|q(p)\rangle_{\rm tree}, 
$
where $\Gamma$ denotes Dirac matrices and $p$ is the off-shell
quark momentum.
In non-perturbative renormalization on the lattice, the matrix element
$\langle q(p)|{\cal O}_\Gamma|q(p)\rangle $ is constructed
with the vertex and wave functions obtained from the Fourier 
transformed quark Green functions in Landau gauge.
This method, which we refer to as NPMOM method hereafter, 
is expected to work if we can find
the appropriate region for the external quark momentum
$\Lambda_{\rm QCD}\ll p\ll a^{-1}$ 
to avoid the
non-perturbative contributions and the cut-off effects simultaneously.
 
The NPMOM method has been actively being pursued for various
operators in different 
actions\cite{npmom_gockeler,bk_w_jlqcd,npmom_qm_ks,npmom_4fermi,dwf_rbc}.
The studies, however, reveal that it is not a trivial task to
find the appropriate region of $p$: it depends on each operator
strongly; moreover, it is hard to find it
for some operators.
At the present stage it is not clear to what extent 
the non-perturbative contaminations and the cut-off effects
can be controlled quantitatively. 
Another concern is the issue of Gribov copies in the Landau gauge 
fixing. We have no definite way to estimate 
the ambiguities induced by the choice of Gribov copies.



The technical difficulties in the NPMOM method can be entirely
overcome by another non-perturbative method that uses the
Schr{\"o}dinger functional(SF) as the renormalization scheme\cite{sf}. 
The SF scheme is essentially a finite-volume renormalization 
technique. QCD is considered in a finite space-time volume 
of physical size $L$ in all directions, where all renormalized
quantities are defined at scale $\mu=1/L$.
A change of the lattice size at fixed bare parameters
is a change of the renormalization scale.
For the bilinear operator $\bar\psi \Gamma\psi$ 
a possible choice of the renormalization condition is
$
Z_\Gamma(g_0,L/a)=c\sqrt{f_{\rm ext}}/f_\Gamma(t)|_{t=L/2},
$
where 
$f_\Gamma(t)=\int d^3{\vec y}d^3{\vec z}
\langle {\bar \psi_f}({\vec x},t)\Gamma\psi_{f^\prime}({\vec x},t)
{\bar \zeta}_{f^\prime}({\vec y},0)\Gamma\zeta_f({\vec z},0)\rangle$
and  
$f_{\rm ext}\hfill=\hfill\int d^3{\vec y}^\prime d^3{\vec z}^\prime
d^3{\vec y}d^3{\vec z}
\langle {\bar \zeta}_f^\prime({\vec y}^\prime,L)\Gamma
\zeta_{f^\prime}^\prime({\vec z}^\prime,L)\\ 
{\bar \zeta}_{f^\prime}({\vec y},0)\Gamma\zeta_f({\vec z},0)\rangle$
with $f$, $f^\prime$ the flavor.
$\zeta,\dots,{\bar \zeta}^\prime$ denote the boundary quark fields
at $t=0$ and $L$ with zero spatial momentum projection.
The constant $c$ in the renormalization condition 
is properly chosen such that $Z_\Gamma=1$ at tree-level.

There are three main advantages in the SF scheme.
(i) Gauge invariance: It is not necessary to fix the gauge. 
(ii) Mass independent renormalization:
This scheme allows us to perform numerical simulations at zero
quark mass. We can specify the boundary conditions
such that the lowest eigenvalue of the free Dirac operator is 
lifted by a gap of order $1/L$ at vanishing quark mass\cite{sf_sint}. 
(iii) Non-perturbative renormalization group:
The scale evolution of a factor of two for
the renormalization constant $Z(\mu)$, which
is expressed by the step scaling functions
$\sigma({\bar g}^2(L))=Z(2L)/Z(L)$, can be measured
by increasing the lattice size $L$ to $2L$. 
In the next step the lattice spacing $a$ is enlarged by a factor of two 
keeping the renormalized coupling ${\bar g}^2$ fixed such that
the number of lattice sites $L/a$ is reduced by a factor of two. 
Repeating this procedure we can connect the perturbative regime to
the hadronic scale: $Z(L_{\rm max})=Z(2^{-k}L_{\rm max})
\sigma({\bar g}^2(2^{-k}L_{\rm max}))\cdots
\sigma({\bar g}^2(2^{-1}L_{\rm max}))$ for $k\ge 1$.
We emphasize that the step scaling 
function is defined in the SF scheme and should be 
independent of the lattice regularization.  

The ALPHA Collaboration demonstrates the effectiveness
of this method calculating the non-perturbative scale
evolution of the renormalized quark mass with $N_f=0$ 
in the SF scheme,
where the scale evolution from the hadronic scale to the perturbative
regime around 100GeV is described within $1\%$ errors\cite{sf_qmrun}.
Besides the quark mass the SF renormalization method 
has been also applied to
the twist-two operator associated with the non-singlet parton 
density\cite{sf_sf,structuref} and the static-light axial vector current
relevant to the $B$ meson decay constant\cite{sf_kurth}. 


\subsection{$B_K$ with the Wilson quark action}

\begin{figure}[t]
\centering{
\hskip -0.0cm
\psfig{file=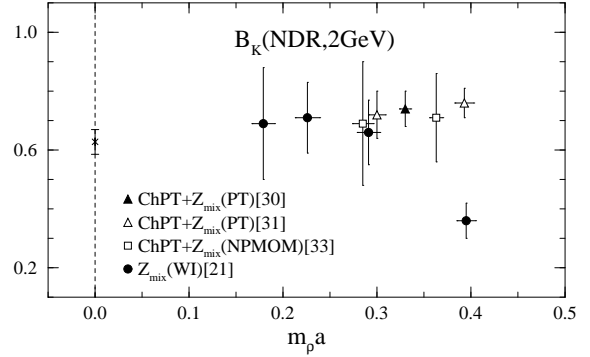,width=75mm,angle=-90}
\vskip -10mm  }
\caption{Quenched $B_K({\rm NDR, 2GeV})$
with the Wilson (filled symbols) and SW (open symbols) 
quark actions  as a function of $m_\rho a$. 
For the KS result in the continuum limit (cross) 
see Fig.~\protect{\ref{fig:bk_ks}}.}
\label{fig:bk_w_all}
\vspace{-8mm}
\end{figure}

\begin{figure}[t]
\centering{
\hskip -0.0cm
\psfig{file=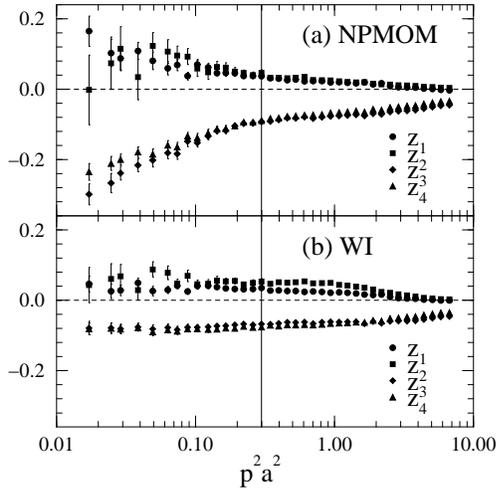,width=65mm,angle=-90}
\vskip -10mm  }
\caption{ Mixing coefficients $z_1,\dots,z_4$
obtained with (a) the NPMOM method and 
(b) the Ward identity method as a function of $p^2a^2$  
around strange quark mass 
with the Wilson quark action at $\beta=6.3$. } 
\label{fig:bk_w_zmix}
\vspace{-8mm}
\end{figure}

We have two purposes to calculate $B_K$ with the Wilson quark action.
One is to demonstrate that the Wilson result is consistent with 
the KS result, which would give confidence
to the lattice QCD calculation of $B_K$.
The other is an application to the heavy-light meson system
for which the interpretation of flavor
quantum numbers with the KS action is problematic.
 
While steady progress has been achieved so far with the KS
quark action, studies with the Wilson quark action are rather
stagnant. This is due to a non-trivial mixing between
the weak four-fermi operator 
${\cal O}_{LL}={\bar s}\gamma_\mu(1-\gamma_5)d
\cdot{\bar s}\gamma_\mu(1-\gamma_5)d$
and other four-fermi operators with different chiralities.
One of the mixing operators is ${\bar s}\gamma_5d\cdot{\bar s}\gamma_5d$.
Numerical studies show that the magnitude of the matrix element
$\langle {\bar K}^0|{\bar s}\gamma_5d\cdot{\bar s}\gamma_5d|K^0\rangle$
is a factor of 10 larger than that of  
$\langle {\bar K}^0|{\cal O}_{LL}|K^0\rangle$ 
at the point of the physical kaon mass
(see, {\it e.g.,} refs.~\cite{bk_w_gupta,bk_w_jlqcd}).
This can be reasonably expected 
from the ratio of their vacuum saturation approximations: 
\be
\frac{11}{16}
\frac{\langle {\bar K}^0|{\bar s}\gamma_5d|0\rangle 
\langle 0|{\bar s}\gamma_5d|K^0\rangle}
{\langle {\bar K}^0|{\bar s}\gamma_\mu\gamma_5d|0\rangle 
\langle 0|{\bar s}\gamma_\mu\gamma_5d|K^0\rangle}
-\frac{1}{16} \approx -17
\ee 
with the aid of the PCAC relation.
In this case the mixing problem is not adequately treated
with one-loop perturbation theory, because two-loop contributions
are potentially large:
$|\alpha_{\rm latt}^2
\langle {\bar K}^0|{\bar s}\gamma_5d\cdot{\bar s}\gamma_5d|K^0\rangle/
\langle {\bar K}^0|{\cal O}_{LL}|K^0\rangle| \\ \simgt 0.1$,
where we assume $\alpha_{\rm latt}=0.15-0.2$ as the currently accessible
strong coupling constant with tadpole improvement.
This fact compels us to control the operator mixing 
non-perturbatively.

Three major non-perturbative methods have been proposed so far.
The most conventional method is to rely on chiral perturbation 
theory\cite{bk_w_gupta,bk_w_ukqcd}.
The mixing structure of the weak four-fermi operator is
expressed by
$
{\hat{\cal O}_{LL}}={\cal O}_{LL}
+\sum_{i=1}^4 z_i {\cal O}_i,
$
where  the mixing operators are arranged into the Fierz eigenbasis 
given by
${\cal O}_1=SS+TT+PP$,
${\cal O}_2=SS-TT/3+PP$,
${\cal O}_3=VV-AA+2(SS-PP)$ and
${\cal O}_4=VV-AA-2(SS-PP)$,
with
$VV={\bar s}\gamma_\mu d \cdot {\bar s}\gamma_\mu d$,
$AA={\bar s}\gamma_\mu\gamma_5 d \cdot {\bar s}\gamma_\mu\gamma_5 d$,
$SS={\bar s}d \cdot {\bar s}d$,
$PP={\bar s}\gamma_5 d \cdot {\bar s}\gamma_5 d$ and
$TT={\bar s}\sigma_{\mu\nu} d \cdot {\bar s}\sigma_{\mu\nu} d/2$.
The authors of ref.~\cite{bk_w_gupta} 
assume the general form of the chiral expansion up to $O(p^4)$
for the matrix elements of ${\hat{\cal O}_{LL}^{\rm pert}}$:
$
\langle {\bar K}^0(p_f)|{\hat{\cal O}_{LL}^{\rm pert}}| K^0(p_i)\rangle
\hfill=\hfill \alpha\hfill +\hfill 
\beta m_K^2\hfill +\hfill \gamma p_i\cdot p_f \\
+\delta_1 m_K^4+\delta_2m_K^2 p_i\cdot p_f +\delta_3(p_i\cdot p_f)^2, \nn
$
where chiral logarithms are ignored. 
One-loop perturbative expressions are used for the mixing coefficients
$z_1,\dots,z_4$ in ${\hat {\cal O}_{LL}^{\rm pert}}$.
The coefficients in the chiral expansion 
of $\langle {\bar K}^0(p_f)|{\hat{\cal O}_{LL}^{\rm pert}}| K^0(p_i)\rangle$
can be determined by changing the quark mass and the
spatial momentum of ${\bar K}^0$ and $K^0$, with which 
we can eliminate pure lattice artifacts of $\alpha$, $\beta$ and
$\delta_1$ terms. In practical applications of this method,
however, it is difficult to fix $\delta_1$, $\delta_2$ and $\delta_3$
with good precision.
The results for $B_K$ are plotted in Fig.~\ref{fig:bk_w_all}(triangles), 
which seem to be reasonable estimates compared
with the KS result at the continuum limit.
Nonetheless, this method is not promising 
for controlling systematic errors, since
it contains unknown uncertainties from higher order effects of 
ChPT which survive even in the continuum limit.
Another defect of this method is that it cannot be 
applied to the heavy-light meson system. 

The second method is to determine the mixing coefficients
$z_1,\dots,z_4$ using the NPMOM method\cite{npmom_4fermi}.
We require the Fourier transformed vertex function
should be proportional to the 
tree-level\hfill Dirac\hfill structure:
\hfill $G_s^{-1}(p)G_s^{-1}(p) \\ \langle {\hat {\cal O}}_{ LL}
[ss{\bar d}{\bar d}](p)\rangle G_{\bar d}^{-1}(p)G_{\bar d}^{-1}(p)
\hfill\propto\hfill  
\gamma_\mu(1\hfill -\hfill \gamma_5)\\ \otimes \gamma_\mu(1-\gamma_5),$
where $[ss{\bar d}{\bar d}](p)$ represents 
$s(p)s(p){\bar d}(p){\bar d}(p)$.
$G_{s,{\bar d}}(p)$ denote the free quark propagators 
with off-shell quark momentum $p$.
In Fig.~\ref{fig:bk_w_zmix}(a) we plot the mixing coefficients 
$z_1,\dots,z_4$ as a function of $p^2a^2$, where
the vertical line corresponds to about 2GeV.  
As mentioned in the previous subsection,
this method can work under the condition 
$\Lambda_{\rm QCD}\ll p\ll a^{-1}$. 
A strong $p$ dependence of $z_1,\dots,z_4$ below $p\approx 2$GeV 
indicates that
non-perturbative contamination plagues the mixing coefficients
up to around 2GeV.
The Gribov uncertainty in 
Landau gauge fixing is another source of systematic errors.

In the third method the mixing coefficients $z_1,\dots,z_4$
are determined to satisfy the continuum Ward identity
up to $O(a)$ for the external off-shell quarks 
with momentum $p$\cite{bk_w_jlqcd,chwi_w}:
$
\langle\nabla_\mu A_\mu{\hat {\cal O}}_{LL} 
[ss{\bar d}{\bar d}](p)\rangle
=2m_q\langle P{\hat {\cal O}_{LL}} 
[ss{\bar d}{\bar d}](p)\rangle
+\langle i\delta_A\{{\hat {\cal O}_{LL}} 
[ss{\bar d}{\bar d}](p)\}\rangle +O(a).
$
It should be stressed that this method can fix the mixing
coefficients without $O(m_q)$ ambiguities for 
the $(27,1)$-operator in the flavor 
SU(3)$_L\times$SU(3)$_R$ representation
which is relevant to $B_K$ and $K\rightarrow\pi\pi[I=2]$ amplitude,
while some ambiguities of $O(m_q)$ remain 
in the case of the $(8,1)$-operator associated with
$K\rightarrow\pi\pi[I=0]$ decay\cite{chwi_w}.
Figure~\ref{fig:bk_w_zmix}(b) shows the momentum
dependence of the mixing coefficients.
The Ward identity method requires the condition $p\ll a^{-1}$ to avoid
the cut-off effects. 
We observe only weak dependence over a wide range
$p^2 a^2\simlt 1.0$, which contrasts to the  
NPMOM results in Fig.~\ref{fig:bk_w_zmix}(a).
In principle, however, both results are expected to be consistent
at the limit of large external quark momentum\cite{npmom_4fermi}.
The Gribov uncertainties in this method could be avoided 
by employing the Schr{\"o}dinger functional technique.

In Fig.~\ref{fig:bk_w_all} we summarize recent quenched 
results for $B_K$ with the Wilson and SW quark actions.
It is encouraging to be able to deal with the mixing problem
without ChPT; still,
large errors hinder the direct comparison with 
the KS result in the continuum limit.


\section{$K\rightarrow \pi\pi$ decay amplitudes}

There are two major difficulties in the lattice QCD calculation
of $K\rightarrow \pi\pi$ decay amplitudes.
The first one is the phase shift in $\pi$-$\pi$ final state interactions,
which cannot be directly calculated with real Green functions
on a Euclidean lattice\cite{kpipi_mt}. Secondly, we have to control
the operator mixing. For the Wilson quark action we have
$
{\cal O}_{\pm}^{\rm cont}=z_{\pm}{\cal O}_{\pm}^{\rm latt}
+z_{\pm}^{\prime}{\cal O}_{\pm}^{\prime}
+(m_c-m_u)\{ z_3 {\bar s}d+z_3^{\prime}(m_s-m_d){\bar s}\gamma_5 d
+z_5 g_0 {\bar s}\sigma_{\mu\nu} F_{\mu\nu}d\},
$
where ${\cal O}_{\pm}^{\prime}$ are dimension-six operators arising
from the explicit chiral symmetry breaking in the action.
The mixing with ${\bar s}d$, ${\bar s}\gamma_5 d$ and 
$g_0 {\bar s}\sigma_{\mu\nu} F_{\mu\nu}d$ is generated by
the ``eye'' diagram, in which two quarks in ${\cal O}_{\pm}$
generate a quark loop by contracting with each other. 
The factors $(m_c-m_u)$ and $(m_s-m_d)$ are required by GIM mechanism
and CPS symmetry\cite{cps} respectively.

The simplest solution to overcome these difficulties
is calculating the amplitude 
$\langle \pi({\vec p}=0)\pi({\vec p}=0)|
{\cal O}_{\pm}^{\rm cont}|K({\vec p}_K=0)\rangle$
with $m_s=m_d$\cite{kpipi_bernard}. 
The final state pions at rest relative to each other
generate no phase shift; the operator mixing is completely avoided
by CPS symmetry and parity conservation.
This method, however, requires ChPT
to convert the amplitudes with the unphysical condition to
those with the physical one.
The next simplest method may be evaluation of 
the $K\rightarrow \pi$ matrix
elements with the KS quark action\cite{cps,ks_ks}.
Although the mixing with ${\cal O}_{\pm}^{\prime}$ can be avoided
by the $\rm U_A(1)$ chiral symmetry retained in the action,
subtraction of the lower dimensional operator is still necessary.
This method also has to rely on ChPT. 
A new idea to avoid the operator mixing problem
was recently proposed\cite{kpipi_ope},
which is 
an attempt to directly calculate the hadronic matrix element of the 
$T$-product of weak currents $T[J_{\mu L}(x)J^\dagger _{\mu L}(0)]$  
in the region $ a  \ll |x| \ll \Lambda_{\rm QCD}^{-1}$. 
However, its practical feasibility is questionable even 
with 1 Teraflops machine because of the requirement of
large cut-off. 

In recent years Golterman and his collaborators\cite{kpipi_gl,kpipi_gp} 
have studied
the $K\rightarrow\pi\pi$ decay amplitudes using 
ChPT and the quenched formulation of ChPT(QChPT)\cite{qchpt} up to one-loop
level, which enable us to investigate
three systematic effects contaminating the lattice QCD calculation:
unphysical kinematics, finite volume effects, and quenching.
Only the first one can be estimated at the tree level.
Although an $O(m_K^2/\Lambda_\chi^2)$ uncertainty still
remains in this calculation  due to the unavailability
of the $O(p^4)$ low-energy constant terms, we should note two points:
First, finite volume effects can arise only from the loop diagrams.
Second, we can investigate the difference 
of the chiral properties between 
full QCD and quenched QCD, which emerge at the one-loop level.

\begin{figure}[t]
\centering{
\hskip -0.0cm
\psfig{file=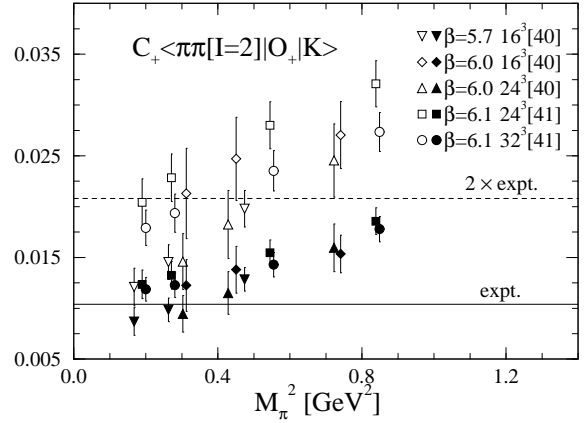,width=75mm,angle=-90}
\vskip -10mm  }
\caption{$K\rightarrow\pi\pi$ decay amplitude in $\Delta I=3/2$
channel with (filled symbols) and without (open symbols) 
one-loop corrections of QChPT. Data are plotted as a function of
lattice meson mass.}
\label{fig:kpipi_i2}
\vspace{-8mm}
\end{figure}

In Fig.~\ref{fig:kpipi_i2} we plot the physical
$K\rightarrow\pi\pi[I=2]$ amplitude as a function of the degenerate 
$K$ and $\pi$ meson mass on the quenched lattice.
They are obtained from the matrix element 
$\langle \pi({\vec p}=0)\pi({\vec p}=0)|
{\cal O}_{+}^{\rm cont}|K({\vec p}_K=0)\rangle$
calculated with the Wilson quark action\cite{kpipi_tasi,kpipi_jlqcd}, 
where ${\cal O}_{+}^{\rm cont}$ is renormalized in NDR scheme
at the scale 2GeV employing the tadpole improvement.  
The conversion to the physical matrix element is made by
\be
\langle \pi\pi|{\cal O}_{+}|K\rangle_{\rm phys}
=\frac{m_K^2-m_\pi^2}{2M_\pi^2}\cdot Y\cdot
\langle\pi\pi|{\cal O}_{+}|K\rangle_{\rm latt},
\label{eq:kpipi_oneloop}
\ee
where the factor $Y$ denotes the one-loop contribution given by
\be
Y=\frac{1+\frac{m_\pi^2}{(4\pi f_\pi)^2}[-104.73-29.57{\rm log}
\left(\frac{m_\pi}{\Lambda}\right)^2]}
{1+\frac{M_\pi^2}{(4\pi F_\pi)^2}[-3{\rm log}
\left(\frac{M_\pi}{\Lambda}\right)^2+F(M_\pi L)]}.
\label{eq:kpipi_y}
\ee
$F(M_\pi L)$ represents finite size effects, 
whose expression is $F(M_\pi L)=17.827/(M_\pi L)+12\pi^2/(M_\pi L)^3$.
The tree-level results(open symbols), which are 
obtained by setting $m_\pi=136$MeV, $m_K=497$MeV and 
$Y=1$ in eq.(\ref{eq:kpipi_oneloop}),
are about two times larger 
than the experimental value.
However, once we include the one-loop effects with $Y$ 
of eq.(\ref{eq:kpipi_y}) employing
$f_\pi=F_\pi=132$MeV and $\Lambda=1$GeV,
the lattice results are reduced and are
much closer to the experimental value.  
The one-loop corrections are rather substantial.
For example, the corrections are $-40\%$ for unphysical 
kinematics,  $-30\%$ for
finite volume effects, and $+30\%$ for quenching
around $M_\pi\approx m_K$ at $\beta=6.1$ 
with the $24^3$ spatial volume(squares). 
It is also noted that
the size dependence observed in the tree-level results
is almost explained through the one-loop calculation.

For the $\Delta I=1/2$ amplitude,
the one-loop calculation with QChPT\cite{kpipi_gp} 
finds that the quenched matrix element 
$\langle \pi({\vec p}=0)\pi({\vec p}=0)|
{\cal O}_{\pm}^{\rm cont}|K({\vec p}_K=0)\rangle$
has finite volume contributions
in proportion to $M_\pi L$, which arise from the  
rescattering diagram of the two pions.  
Although this fact could jeopardize the
numerical calculation of 
$\langle \pi({\vec p}=0)\pi({\vec p}=0)|
{\cal O}_{\pm}^{\rm cont}|K({\vec p}_K=0)\rangle$
in the quenched approximation,
fortunately, we have had no work so far.
This dangerous finite size effect caused by the
$\pi$-$\pi$ interactions can be avoided by
employing the  $K\rightarrow \pi$ matrix element.
Its numerical calculation was tried by the OSU group
employing the KS quark action\cite{kpipi_i0}.
Although their result shows
a consistency with the experimental value 
around $a^{-1}\approx 2.8$GeV,
it is not convincing because of their use of the tree-level ChPT
and the statistical error of $40\%$. 


\section{Proton decay amplitudes}

\begin{figure}[t]
\centering{
\hskip -0.0cm
\psfig{file=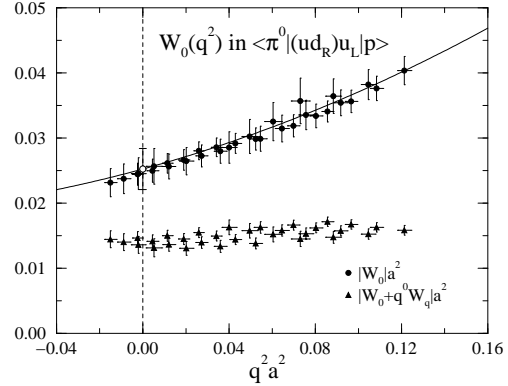,width=65mm,angle=-90}
\vskip -10mm  }
\caption{$q^2a^2$ dependence of the relevant form factor (circles) in
$\langle \pi^0|(ud_R)u_L|p\rangle$, together with
the form factor (triangles) employed in ref.~\protect{\cite{pd_gavela}} for
comparison. Solid line denotes the fitting result.} 
\label{fig:pd_qfit}
\vspace{-8mm}
\end{figure}


One of the most important features of the study of 
the baryon number violating processes is that 
the low energy effective theory is described
in terms of  SU(3)$\times$SU(2)$\times$U(1) symmetry,
which enables us to make a model independent analysis.
All dimension-six operators associated with
baryon number violating processes are classified
in the four types under the
requirement of SU(3)$\times$SU(2)$\times$U(1) invariance 
\cite{pd_wwz,pd_aw}.
In the two-component notation of ref.~\cite{pd_aw}
they are given by
${\cal O}^{(1)}_{abcd}=
(d^{\alpha}_{aR}u^{\beta}_{bR})
(q^{\gamma}_{icL}l_{jdL})
\epsilon_{\alpha\beta\gamma}\epsilon_{ij}$,  
${\cal O}^{(2)}_{abcd}=
(q^{\alpha}_{iaL}q^{\beta}_{jbL})
(u^{\gamma}_{cR}l_{dR})
\epsilon_{\alpha\beta\gamma}\epsilon_{ij}$,  
${\cal O}^{(3)}_{abcd}=
(q^{\alpha}_{iaL}q^{\beta}_{jbL})
(q^{\gamma}_{kcL}l_{ndL})
\epsilon_{\alpha\beta\gamma}\epsilon_{in}\epsilon_{jk}$ and 
${\cal O}^{(4)}_{abcd}=
(d^{\alpha}_{aR}u^{\beta}_{bR})
(u^{\gamma}_{cR}l_{dR})
\epsilon_{\alpha\beta\gamma}$,
where $\alpha$,$\beta$,$\gamma$ are SU(3) color indices, 
$i$,$j$,$k$,$n$ are SU(2) indices, $a$,$b$,$c$,$d$ are generation
indices and $L$ and $R$ denote left- and right-handed fields.
Specifying the decay processes of interest,
in our case (proton,neutron)$\rightarrow$($\pi$,$K$)
+($\bar \nu$, $e^+$, $\mu^+$),
we can list the complete set of independent matrix elements 
in QCD with the assumption of isospin symmetry:
$
\langle \pi^0|\epsilon_{\alpha\beta\gamma}
(u^{\alpha}d^{\beta}_{R,L}) u^{\gamma}_{L}|p\rangle$, 
$\langle \pi^+|\epsilon_{\alpha\beta\gamma}
(u^{\alpha}d^{\beta}_{R,L}) d^{\gamma}_{L}|p\rangle$, 
$\langle K^0|\epsilon_{\alpha\beta\gamma}
(u^{\alpha}s^{\beta}_{R,L}) u^{\gamma}_{L}|p\rangle$, 
$\langle K^+|\epsilon_{\alpha\beta\gamma}
(u^{\alpha}d^{\beta}_{R,L}) s^{\gamma}_{L}|p\rangle$, 
$\langle K^+|\epsilon_{\alpha\beta\gamma}
(u^{\alpha}s^{\beta}_{R,L}) d^{\gamma}_{L}|p\rangle$ and
$\langle K^+|\epsilon_{\alpha\beta\gamma}
(d^{\alpha}s^{\beta}_{R,L}) u^{\gamma}_{L}|p\rangle$. 
All we have to calculate in lattice QCD are these matrix elements.

Under the requirement of Lorentz invariance,
we find that
the matrix elements of ${\cal O}^{\slal{B}{}}$ between the nucleon($N$) and
the pseudoscalar(PS) meson can have two form factors:
\be
\langle PS(\vec{p})|{\cal O}^{\slal{B}{}}_L|N^{s}(\vec{k})\rangle
=P_L(W_0(q^2)+W_q(q^2){\slas{q}{}}) u^{s},
\label{eq:pd_sf}
\ee
where $u^s$ denotes the Dirac spinor with spin state $s$.
In the lattice calculation, $\vec{k}=\vec{0}$ is chosen for 
the nucleon spatial momentum and $\vec{p}=\vec{k}-\vec{q}\neq \vec{0}$ for
the PS meson.
Although the $W_q$ term in  eq.(\ref{eq:pd_sf}) is irrelevant 
in the physical decay amplitude, because its contribution 
is $O(m_l)$ after the multiplication of anti-lepton spinor,
we have to disentangle these two form factors
in the lattice QCD calculation.
As pointed out by JLQCD\cite{pd_jlqcd_99}, 
the previous papers\cite{pd_gavela,pd_jlqcd_98} 
gave wrong results without realizing the existence of $W_q$ term:
they eventually evaluated $W_0+q^0W_q$ instead of $W_0$.

\begin{figure}[t]
\centering{
\hskip -0.0cm
\psfig{file=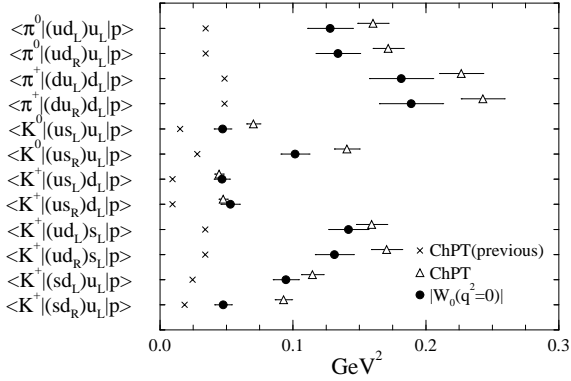,width=75mm,angle=-90}
\vskip -10mm  }
\caption{Comparison of relevant form factors (circle)
with predictions of tree-level ChPT (triangles). 
For the previous ChPT results (crosses) see text.}
\label{fig:pd_all}
\vspace{-8mm}
\end{figure}

In Fig.~\ref{fig:pd_qfit} the quenched results for
$W_0(q^2)$ in $\langle \pi^0|(ud_R) u_{L}|p\rangle$
calculated by JLQCD\cite{pd_jlqcd_99} with the Wilson quark action
at $\beta=6.0$ are compared
with $W_0+q^0W_q$ obtained by following 
the method in ref.~\cite{pd_gavela}.
The magnitude of $W_0(q^2)$ is more than two times larger than 
that of $W_0+q^0W_q$.  
The value at $q^2=0$(open circle) is obtained by fitting
the data with the function of $c_0+c_1 q^2+c_2 q^4 +c_3 m_{ud} +c_4 m_s$.
In Fig.~\ref{fig:pd_all} we compare  
the results obtained by the direct method(circles) with those by 
the indirect one(triangles) using tree-level ChPT, where the 
so-called $\alpha,\beta$ parameters are $|\alpha|=0.015(1){\rm GeV}^3$
and  $|\beta|=0.014(1){\rm GeV}^3$.
We observe that both results are roughly comparable, which
could leads to the conclusion that the large discrepancy between
two methods found in refs.\cite{pd_gavela,pd_jlqcd_98} is mainly due to 
the neglect of the $W_q(q^2)$ term in 
eq.(\ref{eq:pd_sf}). 
Another important point in Fig.~\ref{fig:pd_all} is that 
the JLQCD results are much larger than the tree-level ChPT
predictions with $|\alpha|=|\beta|=0.003$GeV$^3$(crosses), which is 
the smallest value among various model estimations.


\section{Domain wall QCD}
\label{sec:dwf}

If we can make the length of the fifth dimension $N$
sufficiently large
in the practical implementation of the domain wall QCD on the lattice,
the desirable features would emerge as:  
(i) no fine tuning for the chiral limit up to $O(1/a{\rm e}^{-\alpha N})$,
(ii) $O(a^2)$ scaling violation 
up to $O(a{\rm e}^{-\alpha^\prime N})$ and
(iii) no mixing for four-fermi operators 
up to $O(g^2{\rm e}^{-\alpha^{\prime\prime} N})$.
The flavor symmetry is retained at any number $N$.
The KS quark action, however, can 
realize the above features without corrections except
the flavor symmetry.
The cost of numerical studies with the domain wall quark
are the fifth dimension with the extension $N$
and choosing the optimal value for the domain wall height $M$.
The RIKEN-BNL-Columbia group presented
numerical studies for weak matrix elements
relevant to ${\bar K}^0$-$K^0$ mixing, $K\rightarrow\pi\pi$ decays and
the prediction of $\epsilon^\prime/\epsilon$\cite{dwf_rbc},
assuming the naive chiral properties realized 
at the infinite $N$\cite{dwf_rbc}. 
In Fig.~\ref{fig:bk_ks} we already plotted the quenched 
results for $B_K$
by the RIKEN-BNL-Columbia group. Their results are considerably smaller than
the KS results at finite lattice spacings, and the discrepancy
is likely to survive even in the continuum limit. 
At present it is not clear what causes this difference.

\begin{figure}[t]
\centering{
\hskip -0.0cm
\psfig{file=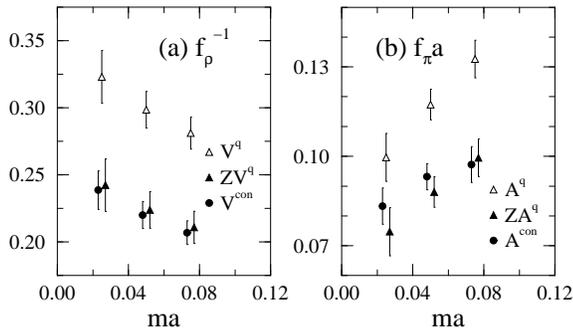,width=75mm,angle=-90}
\vskip -10mm  }
\caption{$f_\rho^{-1}$ and $f_\pi$ obtained from the 
five-dimensional conserved
current (circles) and the four-dimensional 
local current with (filled triangles) and without (open triangles) 
perturbative renormalization factors.}
\label{fig:dwf_fva}
\vspace{-8mm}
\end{figure}

Perturbative calculation in domain wall QCD, which was
initiated in ref.\cite{dwf_pt_at}, has been applied to evaluation of the
renormalization factors for the quark mass\cite{dwf_pt_2,dwf_pt_bsw} and 
the bilinear operators\cite{dwf_pt_2} consisting of four-dimensional quarks
which were referred to as $q(x)$ in ref.~\cite{dwf_fs}.
The Tsukuba group\cite{dwf_pt_2} shows that
$Z_S=Z_P=Z_m^{-1}$ and $Z_V=Z_A$ at $N\rightarrow \infty$,
which is expected when the chiral Ward-Takahashi identity 
hold exactly.
The peculiar feature in the renormalization of the domain wall QCD 
is an appearance of the overlap factor $(1-|1-M|^2)Z_w$
for the four-dimensional quark fields. 
The one-loop coefficient of $Z_w$ 
is of $O(10-10^2)$ for $|1-M|\simgt 0.1$
without mean-field improvement.
This could be dangerous because current numerical studies employ
$M \simgt 1.6$\cite{dwf_rbc}.
In the mean-field analysis, however, they find that $M$ should be
replaced by ${\tilde M}=M+4(u-1)$, where $u^4$ corresponds to the 
expectation value of the plaquette, and the magnitude of 
the one-loop coefficient for $Z_w$ is reduced to 
$O(1)$ for $|1-{\tilde M}|\simlt 0.8$.
This fact suggests that the mean field
improvement is indispensable for the perturbative renormalization factors 
in domain wall QCD.  
The $\rho$ meson decay constant $f_\rho$
obtained with $V^q={\bar q}(x)\gamma_\mu q(x)$
and with the five-dimensional conserved current $V^{\rm con}$ gives
a testing ground for the validity of perturbation theory\cite{dwf_fva}. 
Both results at $\beta=6.0$ are compared in Fig.~\ref{fig:dwf_fva}(a).
They show a consistency within error bars
once the perturbative  
corrections are applied to $V^q$.
We also observe a similar situation for the case of the pion decay constant 
in Fig.~\ref{fig:dwf_fva}(b).
This is an encouraging result; still, the scaling violation effects
should be checked.
The Tsukuba group also calculated the renormalization factors for the three-
and four-quark operators 
associated with proton decay and  
weak interactions\cite{dwf_pt_34}. 
They find that the operators in domain wall QCD
at infinite $N$ can be renormalized without any operator mixing 
between different chiralities, as opposed to the Wilson case.

The good chiral properties of the four-dimensional 
quarks in domain wall QCD 
observed in the numerical simulations and the perturbative
calculations can be understood from the point of view of the 
exact chiral symmetry based on the Ginsparg-Wilson relation.
The partition function of 
domain wall fermion with 
the subtraction of the Pauli-Villars field at finite $N$ 
can be written as a determinant of  
the truncated overlap Dirac operator $D_N$\cite{dwf_tol} 
which satisfies the Ginsparg-Wilson relation 
in the limit of infinite $N$.
The authors of ref.~\cite{dwf_kn} find that the effective Dirac operator
$D_N^{\rm eff}$ for  the four-dimensional quarks 
in domain wall QCD with finite $N$, which is obtained by
integrating out the $N-1$ heavy fermion fields, 
is related to $D_N$:
${D_N^{\rm eff}}^{-1}(x,y)+\delta(x,y)={D_N}^{-1}(x,y)$.
With the use of this relation they show that the Green functions
consisting of $q(x)$ and ${\bar q}(x)$ in domain wall QCD
with finite $N$ can be related to those consisting 
of the fermion fields described by 
the truncated overlap Dirac operator.



\section{Conclusions}

The sizable magnitude of the $O(\alpha^2)$ uncertainties found in
the systematic study of $B_K$ with the KS quark action
makes manifest the necessity of using
the non-perturbative renormalization method.
The essential problem in the non-perturbative renormalization of
QCD, which is the question how to relate the high-energy 
perturbative regime to the low-energy hadronic scale, is
overcome by the Schr{\"o}dinger functional method.
For the Wilson $B_K$ the operator mixing problem 
is now managed non-perturbatively without the use of 
any effective theories; 
while we still need to reduce statistical errors. 

In $K\rightarrow\pi\pi$ studies one-loop corrections
in QChPT can almost explain
the discrepancy between quenched lattice results 
and the experimental value for the $\Delta I=3/2$ transition.

Essential progress has been made 
in evaluating the proton decay amplitudes.
The first correct calculation 
shows that the values of the amplitudes are comparable with the
tree-level ChPT predictions.  

The application of the domain wall quark formulation to the
calculation of the weak matrix elements is a fascinating issue.
Pilot numerical studies have been pursued and
perturbative renormalization factors for the relevant operators 
are now available.

\section{Acknowledgments}

I would like to thank S.~Aoki, C.~Bernard, T.~Blum, M.~Golterman,
N.~Ishizuka, T.~Izubuchi, Y.~Kikukawa, E.~Pallante, Y.~Taniguchi, 
N.~Tsutsui, R.~Sommer, A.~Soni and P.~Weisz
for communicating their results and for discussions.
I would also like to thank C.~Bernard and M.~Golterman 
for comments on the manuscript.
This work is supported in part by a JSPS Fellowship.

\end{document}